\documentclass[11pt,twoside]{article}

\usepackage{asp2004}
\usepackage{epsf}
\usepackage{psfig}
\usepackage{lscape}
\usepackage{amsbsy}

\def\apj{ApJ}
\def\apjs{ApJS}

\def\apjl{ApJL}
\def\jfm{J. Fluid Mech.}

\def\pha{Physica A}
\def\phf{Phys. Fluids}
\def\pre{Phys. Rev. E}
\def\prl{Phys. Rev. Lett.}
\def\aap{A\&A}

\def\mnras{MNRAS}
\def\jcp{J. Comp. Phys.}

\begin{document}
\title{High resolution simulations of supersonic turbulence in molecular clouds}
\author{Alexei G. Kritsuk, Rick Wagner, Michael L. Norman, and Paolo Padoan}
\affil{Department of Physics 
and Center for Astrophysics and Space Sciences,
University of California, San Diego, 9500 Gilman Drive, La Jolla, CA 
92093-0424}

\begin{abstract} 
We present the results of three-dimensional simulations of supersonic
Euler turbulence with grid resolutions up to $1024^3$ points. Our numerical
experiments describe nonmagnetized driven turbulent flows with an isothermal
equation of state and an rms Mach number of 6. 
We demonstrate that the inertial range scaling properties of turbulence in this strongly 
compressible regime deviate substantially from a Kolmogorov-like behavior previously
recovered for mildly compressible transonic flows.
\end{abstract}

\section{Introduction}
Understanding the nature of supersonic turbulence is of fundamental importance
in astrophysics and in aeronautical engineering. In the interstellar medium, 
highly compressible turbulence is believed to control star formation in dense 
molecular clouds. A whole class of more {\em terrestrial} applications deals 
with the drag and stability of projectiles traveling through the air at 
hypersonic speeds.

Molecular clouds have a highly inhomogeneous structure and the intensity of 
their internal motions corresponds to an rms Mach number of the order of 20. 
\citet{larson81} has demonstrated that within the range of scales from 0.1~pc 
to 100~pc, the gas density and the velocity dispersion tightly correlate with 
the cloud size.\footnote{See \citet{kaplan.53} for an earlier version of what
is now known as Larson's relations.} Supported by other independent observational 
facts indicating 
scale invariance, these relationships are often interpreted in terms of 
supersonic turbulence with characteristic Reynolds numbers $Re\!\sim\!10^8$. 
Within a wide range of densities around $10^3$~cm$^{-3}$ the gas temperature
remains close to $\!\approx\!10$~K since the thermal equilibration time at these
densities is shorter than a typical hydrodynamic time scale. Thus, an isothermal 
equation of state can be used as a reasonable approximation. While self-gravity, 
magnetic fields, chemistry, cooling and heating, as well as radiative transfer 
should be ultimately accounted for in turbulent models of molecular clouds, we 
focus here specifically on hydrodynamic aspects of the problem. 

Numerical simulations of {\em decaying} supersonic hydrodynamic turbulence with the 
piecewise parabolic method \citep[PPM]{colella.84} in two dimensions were
pioneered by \citet{passot..88}\footnote{See a review on compressible
turbulence by \citet{pouquet..91} for references to earlier works.} and 
then followed up with high resolution 2D and 3D 
simulations by \citet{porter..92a,porter..92b,porter..94,porter..98}. 
\citet{sytine....00} compared the results of PPM Euler computations 
with PPM Navier-Stokes results and showed that Euler simulations agree well with 
the high-$Re$ limit attained in the Navier-Stokes models. The convergence
in a statistical sense as well as the direct comparison of structures in configuration
space indicate the ability of PPM to accurately simulate turbulent flows over a wide
range of scales.
More recently, \citet{porter..02} discussed measures of intermittency in simulated 
{\em driven} transonic flows at Mach numbers of the order unity on grids up to 
$512^3$ points.
\citet{porter...99} review the results of these numerical studies focusing on
the origin and evolution of turbulent structures in physical space as well as on
scaling laws for two-point structure functions. One of the important results
of this fundamental work is the demonstration of compatibility of a Kolmogorov-type
\citep[K41]{kolmogorov41a} spectrum with a {\em mild} gas compressibility at 
transonic Mach numbers.

Since most of the computations discussed above assume a perfect gas equation
of state with the ratio of specific heats $\gamma=7/5$ or $5/3$ and Mach 
numbers generally below 2, the question remains whether this result will still hold
for near isothermal conditions and {\em hypersonic} Mach numbers characteristic of dense 
parts of star forming molecular clouds where the gas compressibility is much higher.
What kind of coherent structures should one expect to see within the inertial range
of scales in highly supersonic isothermal turbulence? Do low-order statistics of turbulence
follow the K41 predictions closely in this regime? How intermittent is the turbulence? 
The interpretation of astronomical data from new surveys of cold ISM and dust in
the Milky Way by {\em Spitzer} and {\em Herschel} satellites requires more detailed
knowledge of these basic properties of supersonic turbulence.

In this paper we report first results from our large-scale numerical simulations
of driven supersonic isothermal turbulence at Mach 6 with PPM and grid resolutions up to 
$1024^3$ points. We solve the Euler equations in a periodic box of linear size $L=1$
with initially uniform density distribution $\rho\equiv1$ and the sound speed $c\equiv1$.
We initialize the simulation on a grid of $512^3$ points with the random velocity field 
$\pmb{u}_0$ that contains only large-scale power within the range of wavenumbers 
$k/k_{min}\in[1,2]$, where $k_{min}=2\pi$, and corresponds to the rms Mach number $M=6$. 

The same velocity field is then used, with an appropriate normalization, as a steady 
random force (acceleration) to keep the total kinetic energy within the box on an 
approximately constant level during the simulation. The random force is isotropic in 
terms of the total specific kinetic energy per dimension, 
$\left<u_{0,x}^2\right>=\left<u_{0,y}^2\right>=\left<u_{0,z}^2\right>$,
but its solenoidal ($\pmb{\nabla}\cdot\pmb{u}_0^s\equiv0$) and
dilatational ($\pmb{\nabla}\times\pmb{u}_0^c\equiv0$)
components are anisotropic since one of the three directions is dominated by the large-scale
compressional modes, while the other two are mostly solenoidal. The distribution of total
specific kinetic energy $E\equiv\frac{1}{2}\int u^2dV=E^s+E^c$ between the solenoidal 
$E^s$ and dilatational $E^c$ components is such that 
$\chi_0\equiv E_0^s/E_0\approx0.6$. The driving field is helical, but the mean helicity
is very low: $\left<h_0\right>\ll\sqrt{\left<h_0^2\right>}$, where the helicity $h$ is 
defined as $h\equiv\pmb{u}\cdot\pmb{\nabla}\times\pmb{u}$.
In compressible flows with an isothermal equation of state the mean helicity 
$\left<h\right>$ is conserved, as in the incompressible case,
since the Ertel's potential vorticity is identically zero \citep{gaffet85}. 

We first ran the simulation on a grid of $512^3$ points from the initial conditions up to 
five dynamical times to stir the gas within the box. The dynamical time is defined as 
$t_d\equiv L/(2M)$. 
Then we doubled the resolution and evolved the simulation for another $5t_d$ on a grid of 
$1024^3$ points. We allow one dynamical time for relaxation at high resolution
to reach a statistical steady state after regridding. The time-average statistics are computed
using 170 snapshots evenly spaced in time over the final segment of $4t_d$.
We use the full set of 170 snapshots to derive the density statistics, since the density field
displays a very high degree of intermittency. This gives us a very large statistical sample,
e.g. $\sim2\,10^{11}$ measurements are available to determine the probability density function 
(PDF) of the gas density. 
The time-average power spectra discussed below are also based on the full data set.
The velocity structure functions are derived from a sample of 20\%
of the snapshots covering the same period of $4t_d$. The corresponding two-point PDFs are built on 
$2-4\times10^9$ pairs per snapshot each, depending on the pair separation.

\section{Results}
\subsubsection{Time evolution of global variables.} 
\begin{figure}
 \centerline{\hbox{ \hspace{0.0in} 
    \epsfxsize=2.2in
    \epsffile{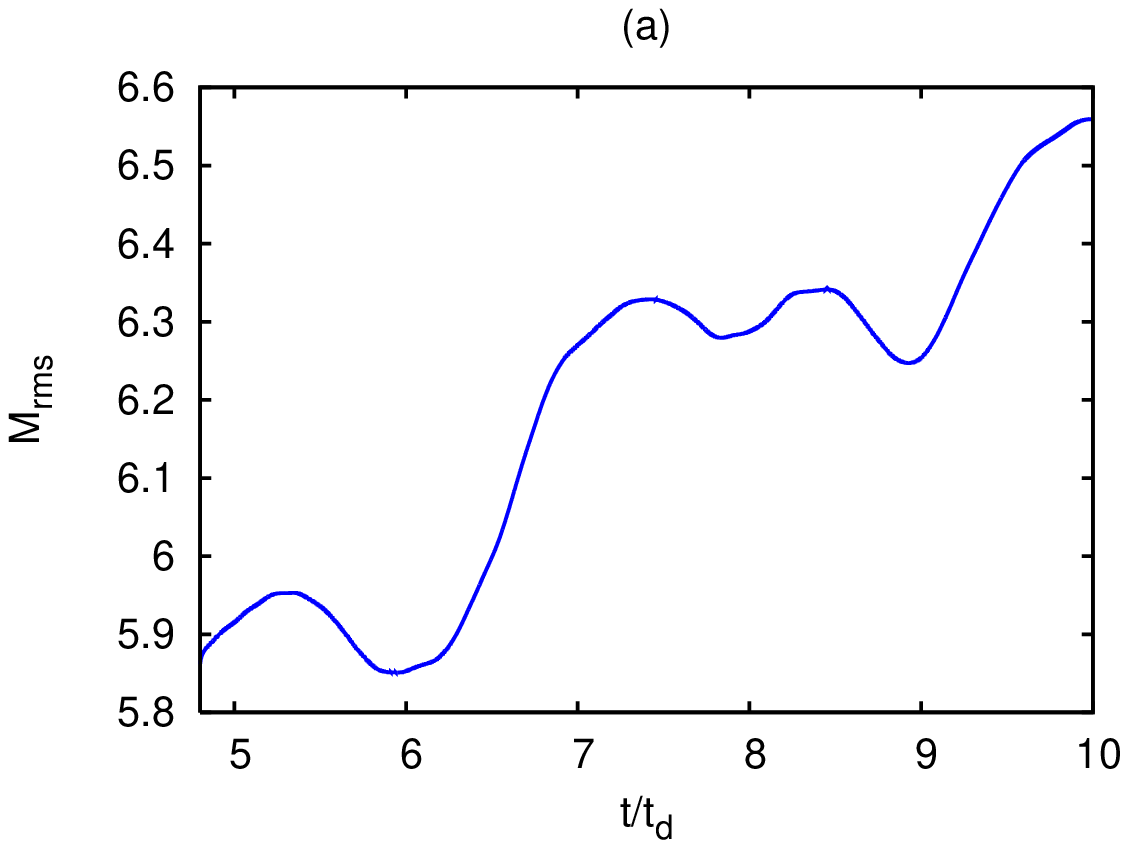}
    \hspace{0.25in}
    \epsfxsize=2.2in
    \epsffile{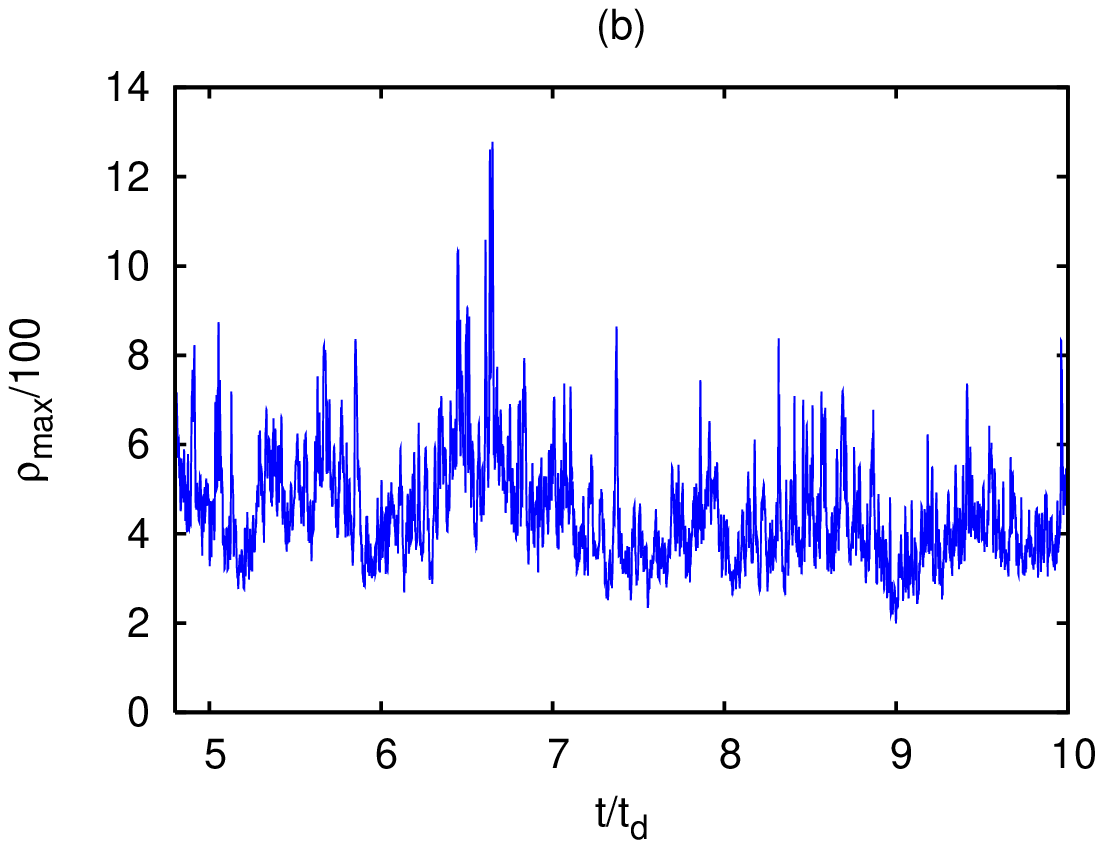}
    }
  }
 \centerline{\hbox{ \hspace{0.0in} 
    \epsfxsize=2.2in
    \epsffile{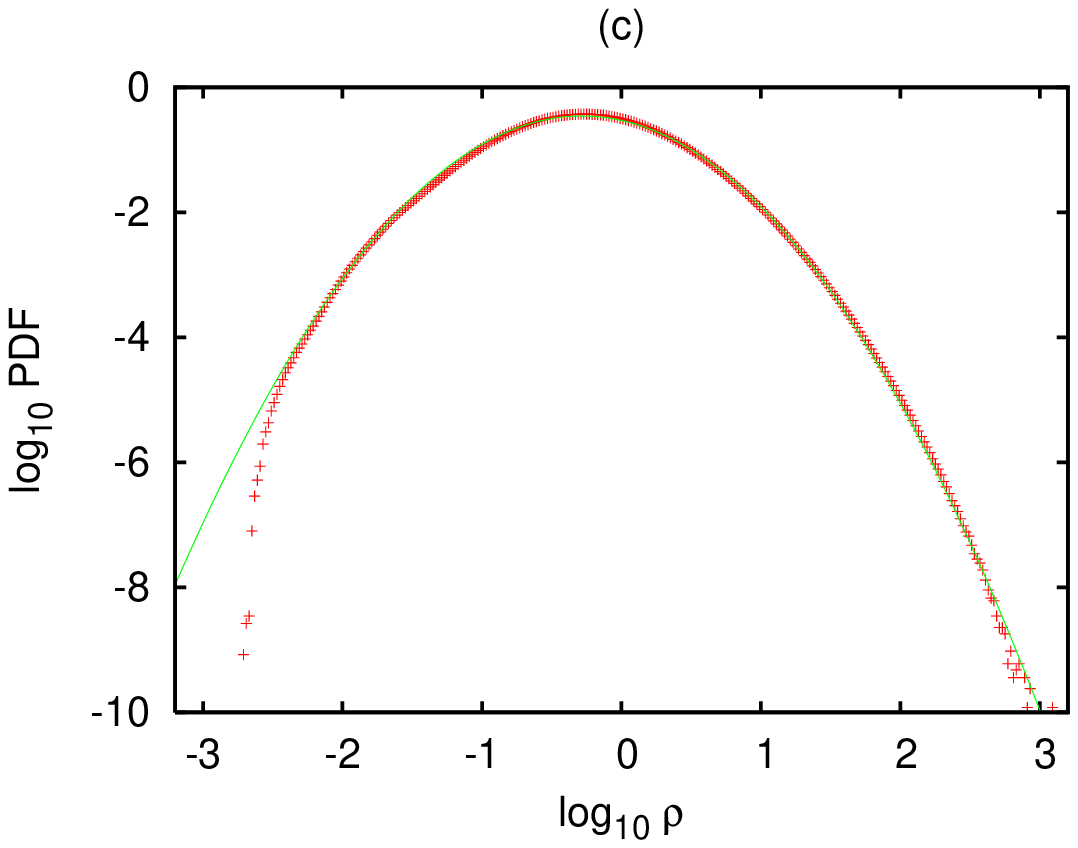}
    \hspace{0.25in}
    \epsfxsize=2.2in
    \epsffile{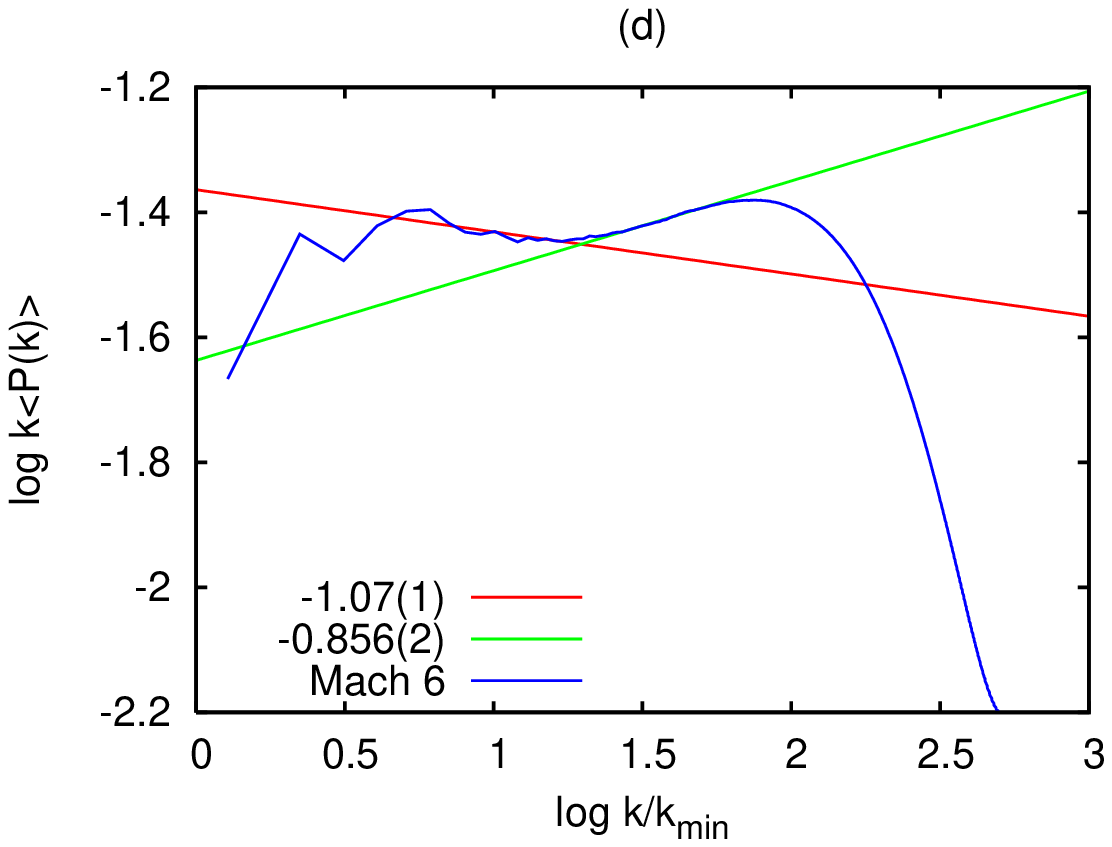}
    }
  }
 \centerline{\hbox{ \hspace{0.0in} 
    \epsfxsize=2.2in
    \epsffile{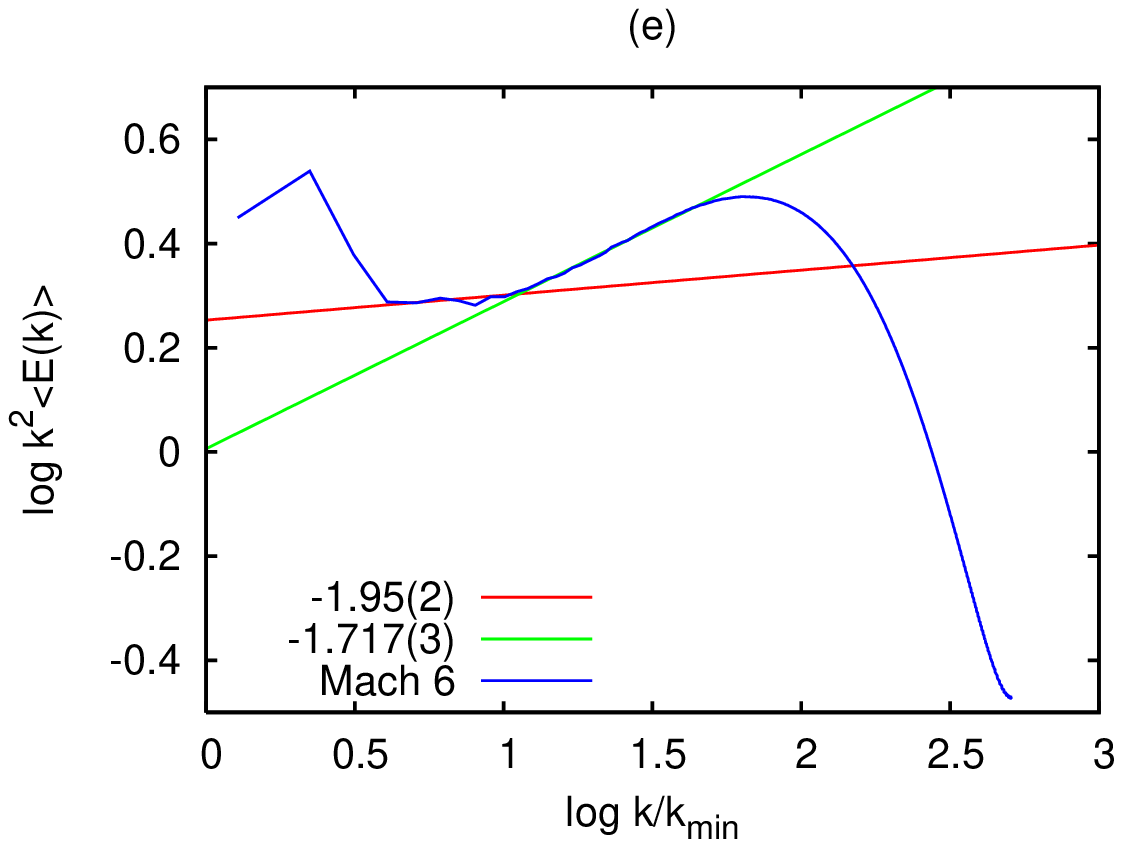}
    \hspace{0.25in}
    \epsfxsize=2.2in
    \epsffile{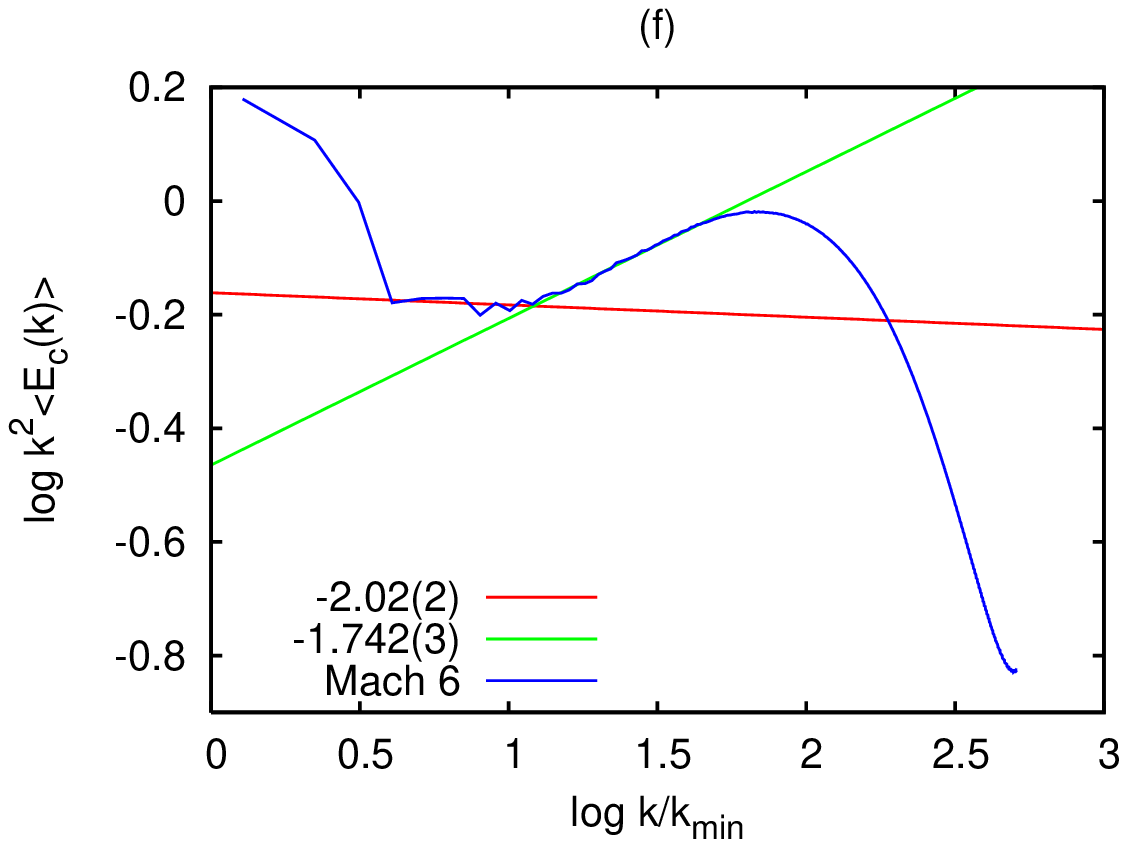}
    }
  }
 \centerline{\hbox{ \hspace{0.0in} 
    \epsfxsize=2.2in
    \epsffile{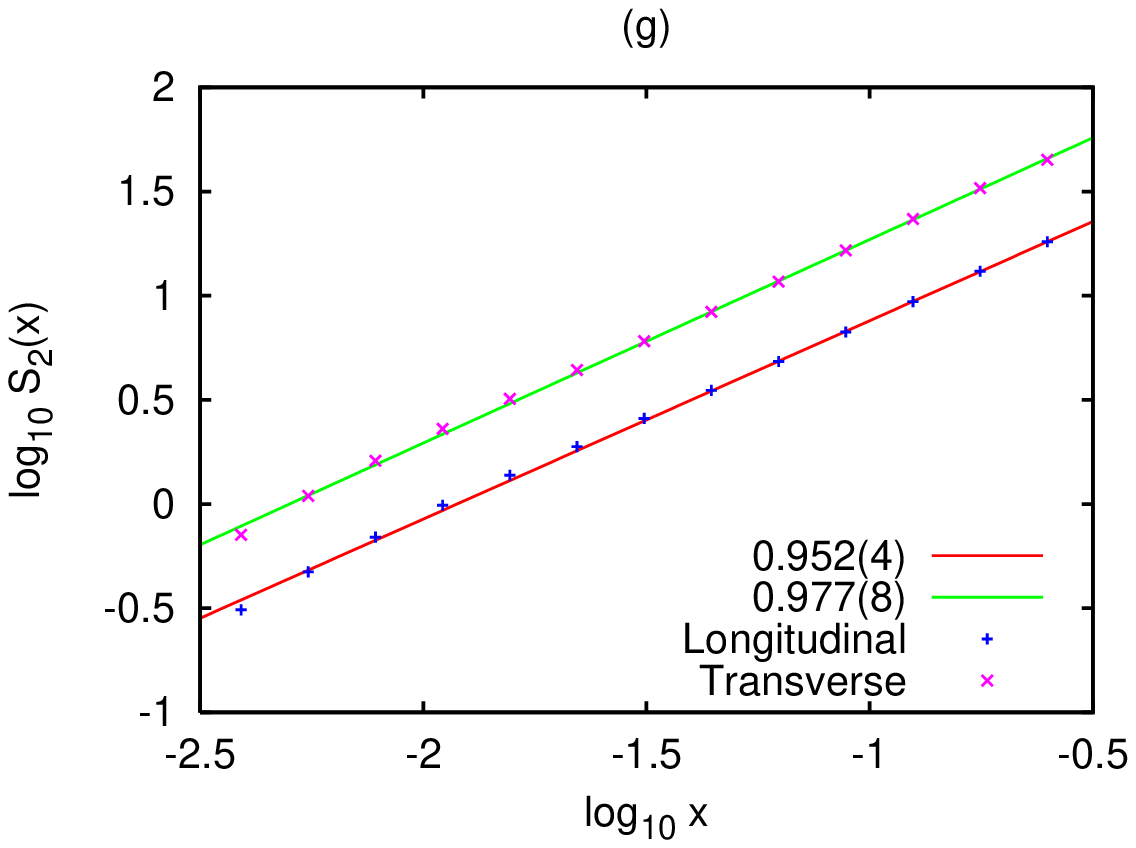}
    \hspace{0.25in}
    \epsfxsize=2.2in
    \epsffile{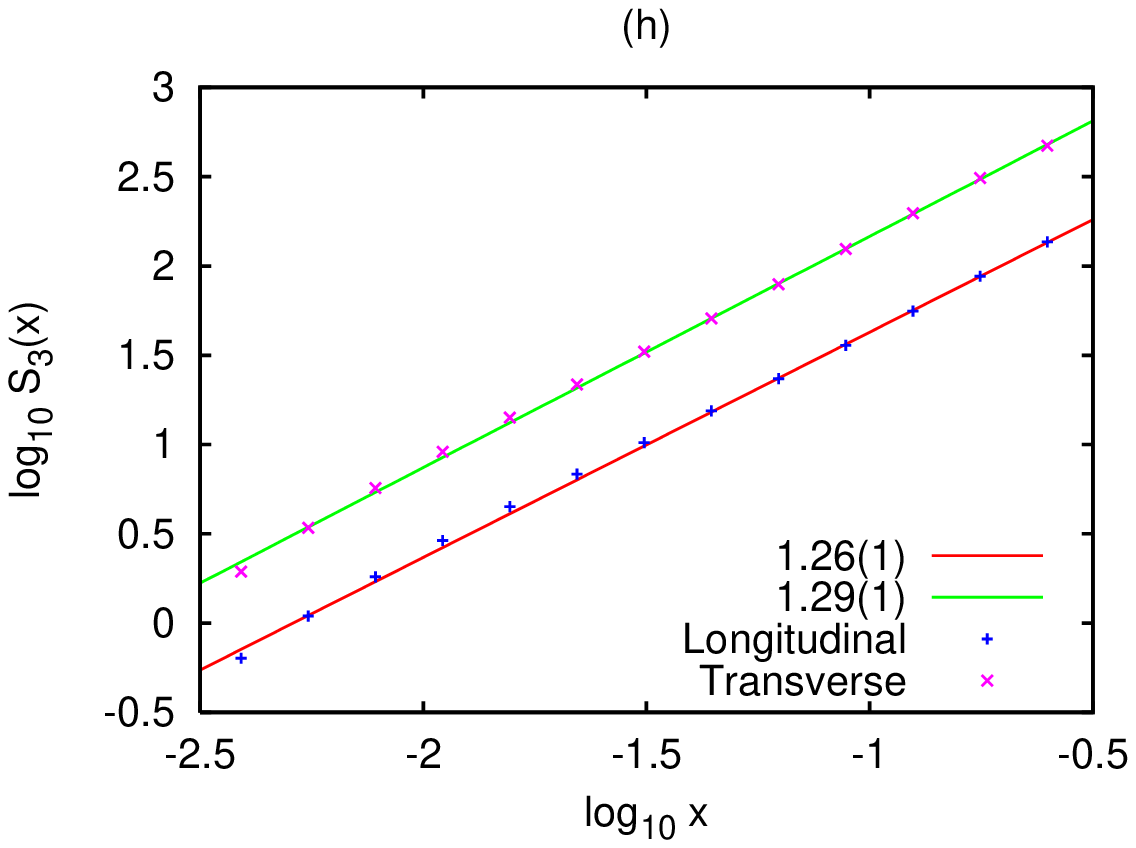}
    }
  }
\caption{Results for the $1024^3$ simulation of driven Mach 6 turbulence with the
ratio of specific heats $\gamma=1.001$:
(a) the time evolution of the rms Mach number; 
(b) the max density as a function of time; 
(c) the histogram for the gas density and the best-fit lognormal; 
(d) the density power spectrum compensated by $k$; 
(e) the velocity power spectrum compensated by $k^2$; 
(f) the power spectrum of dilatational velocity compensated by $k^2$;
(g) the 2nd order structure functions and the best-fit power laws for 
$\log_{10}x\in[-1.5,-0.5]$;
(h) the 3rd order structure functions and the best-fit power laws for 
$\log_{10}x\in[-1.5,-0.5]$.
}
\label{figone}
\end{figure}

The time variations of the rms Mach number
and of the maximum gas density are shown in Fig.~\ref{figone}a, b. The kinetic energy 
oscillates between 18 and 22, roughly following the Mach number evolution. Notice the highly 
intermittent bursts of activity in the plot of $\rho_{max}(t)$.
The time average enstrophy 
$\Omega\equiv\frac{1}{2}\int\left|\pmb{\omega}\right|^2dV\approx10^5$
and the Taylor scale $\lambda\equiv\sqrt{\frac{5E}{\Omega}}\approx0.03=30\Delta$, 
where $\Delta$ is the linear grid spacing and $\pmb{\omega}\equiv\pmb{\nabla}\times\pmb{u}$ 
is the vorticity.
The rms helicity grows by a factor of $7.7$ in the initial phase of the simulation and then
remains roughly constant at a level of $1.2\times10^3$. 
The conservation of the mean helicity is 
satisfied reasonably well as $\left<h\right>$ is contained within $\pm2$\% of its rms value 
during the whole simulation.

\subsubsection{Density PDF.} 
In isothermal turbulence the gas density does not correlate with the
local Mach number. As a result, the density PDF follows a lognormal distribution
\citep{vazquezsemadeni94,padoan..97,passot.98,nordlund.99}. 
Fig.~\ref{figone}c shows our results for the time-average density PDF and its best-fit
lognormal representation. The density distribution is very broad due to the very high 
degree of compressibility in isothermal supersonic conditions. 
The density contrast $\sim\!10^6$ is orders of magnitude higher than in the transonic 
case at $\gamma=1.4$ studied by \citet{porter..02}. 
Notice the excellent quality of the 
fit over more than eight decades in probability and a very low noise level in the data. 
If we express the standard deviation $\sigma$ as a function of Mach number as 
$\sigma^2=\ln(1+b^2M^2)$, we get the best-fit value of $b\approx0.26$ that is smaller 
than $b\approx0.5$ determined in \citet{padoan..97} for supersonic MHD turbulence.
The powerful intermittent bursts in $\rho_{max}(t)$ correspond to large departures
from the time-average PDF caused by head-on collisions of strong shocks. These events are
usually followed by strong rarefactions that are seen as large oscillations in the 
low density wing of the PDF and also in the density power spectrum. Intermittency is 
apparently present in supersonic turbulence. Its signature in turbulence statistics 
will be addressed elsewhere.

\subsubsection{Inertial range scaling.} The power spectrum of the gas density shows
a short straight section with a slope of $-1.07\pm0.01$ in the range of scales from 
$250\Delta$ down to $40\Delta$ followed by flattening due to a power pileup 
at higher wavenumbers, see Fig.~\ref{figone}d.
A similar power-law section, although with a slope of $-1.95\pm0.02$, and an excess of 
power in the same near-dissipation range is also clearly seen in the velocity
power spectrum, Fig.~\ref{figone}e. Apparently, the flattening of the spectra in the
near-dissipation part of the inertial range is due to the so-called bottleneck effect
related to a three-dimensional non-local mechanism of energy transfer between modes of
differing length scales \citep{falkovich94}. The same phenomenon has been observed both 
experimentally and in numerical simulations 
\citep[e.g.][]{porter..94,kaneda....03,dobler...03,haugen.04}.
At the resolution of $512^3$, the bottleneck
effect at high wavenumbers and turbulence forcing at low $k$
leave essentially no room for the uncontaminated inertial range in the $k$-space,
even though the spectrum does show a nice power law. For the density power spectrum,
we would get a slope of $-0.90\pm0.02$ at $512^3$ that is substantially shallower
than $-1.07$ at $1024^3$. Thus, spectral index estimates based on low resolution simulations
\citep[e.g.][]{kim.05} may bear a rather large uncertainty. 
We also note that the time-averaging over many
snapshots is essential to get the correct slope for the density power spectrum since
it exhibits variations on a short (compared to $t_d$) time-scale. The spectrum tends to 
get shallower upon collisions of strong shocks, when the PDF's high density wing rises above 
its average lognormal representation.

The time-average velocity power spectra for the solenoidal and dilatational components
show the inertial range power indices of $-1.92\pm0.02$ and $-2.02\pm0.02$, respectively,
see Fig.~\ref{figone}f for the $k^2$-compensated dilatational spectrum.
Both spectra display flattening due to bottleneck with indices of $-1.70$ and $-1.74$ in 
the near-dissipative range. The fraction of energy in dilatational modes quickly drops 
from about $50$\% at $k=k_{min}$ down to $30$\% at $k/k_{min}\approx50$ and then returns
back to a level of $45$\% at the Nyquist frequency.

Overall, the inertial range scaling of the velocity power spectrum in our highly compressible
turbulence simulations tends to be closer to the Burgers scaling with a power index of $-2$
rather than to the K41 $-5/3$-scaling suggested for the mildly compressible transonic 
simulations by \citet{porter..02}. To substantiate this result, we also computed the scaling
properties of velocity structure functions 
$S_p(x)\equiv\left<\left|\pmb{u}(\pmb{r}+\pmb{x})-\pmb{u}(\pmb{r})\right|^p\right>$ 
of the orders $p=1$, $2$, and $3$, where the averaging is taken for all positions $\pmb{r}$ 
and all orientations of $\pmb{x}$ within the computational domain. Both longitudinal
($\pmb{u}\parallel\pmb{x}$) and transverse ($\pmb{u}\perp\pmb{x}$) structure functions
can be well approximated by power laws $S_p(x)\propto x^{\zeta_p}$, see Fig.~\ref{figone}g
and h. 
The low-order structure 
functions are less susceptible to the bottleneck contamination and might be better
suited for deriving the scaling exponents \citep{dobler...03}.\footnote{Note, however, that
the bottleneck corrections grow with the order $p$ \citep{falkovich94} and influence the
structure functions in a non-local fashion \citep{dobler...03}.} 
The best-fit 2nd order exponents measured for the range of scales between 
$32\Delta$ and $256\Delta$, $\zeta^{\parallel}_2=0.952\pm0.004$ and 
$\zeta^{\perp}_2=0.977\pm0.008$ are substantially 
larger than the K41 predicted value of $2/3$ and agree well with our measured 
velocity power indices. The first order exponents $\zeta^{\parallel}_1=0.533\pm0.002$ and 
$\zeta^{\perp}_1=0.550\pm0.004$ are significantly larger than the K41 index of $1/3$.
We find that the 3rd order scaling exponents $\zeta^{\parallel}_3=1.26\pm0.01$ and 
$\zeta^{\perp}_3=1.29\pm0.01$ are also noticeably off from unity predicted by K41 for the 
incompressible limit and also confirmed by \citet{porter..02} for the transonic regime.
Our measurements for the low order exponents roughly agree with the estimates 
$\zeta^{\perp}_1\approx0.5$, $\zeta^{\perp}_2\approx0.9$, and $\zeta^{\perp}_3\approx1.3$
obtained by \citet{boldyrev..02} for numerical simulations of isothermal Mach 10 MHD 
turbulence at a resolution of $500^3$ points.

The large-scale driving force we used in these simulations is not perfectly isotropic due 
to the uneven distribution of power between the solenoidal and dilatational modes 
(perhaps, a typical situation for the interstellar conditions). We also use a {\em static}
driving force that could potentially cause some anomalies on time-scales of many dynamical
times. However, while strong anisotropies can significantly affect the scaling of high-order 
moments \citep{porter..02,mininni..06}, the departures from Kolmogorov-like scaling we 
observe in the lower order statistics appear to be too strong to be explained solely as a 
result of the specific 
properties of our driving. The sensitivity of our result to turbulence forcing remains to be 
verified with future high resolution simulations involving a variety of driving options.

\section{Conclusions}
Using high-resolution numerical simulations of nonmagnetic highly compressible turbulence 
at rms Mach number of 6, we have demonstrated that scaling exponents of low-order statistics 
deviate substantially from Kolmogorov laws for incompressible turbulence.
A much higher than $1024^3$ resolution is required to possibly trace a transition 
from a steeper supersonic inertial range scaling at lower $k$ to a flatter Kolmogorov-like
transonic scaling at higher wavenumbers.

\acknowledgements
This research was partially supported by a NASA ATP grant NNG056601G,
by NSF grants AST-0507768 and AST-0607675, and by a NRAC allocation MCA098020S.
We utilized computing resources provided by the San Diego Supercomputer Center.


\end{document}